\documentclass{sig-alternate}

\usepackage{booktabs}
\usepackage{multirow}
\usepackage{subfigure}
\usepackage{wrapfig}
\usepackage{url}
\usepackage{hyperref}
\usepackage[utf8]{inputenc}
\usepackage{amssymb}
\usepackage{amsfonts}
\usepackage{amsmath}
\usepackage[usenames,dvipsnames]{xcolor}
\usepackage{tikz}
\usepackage{etoolbox}
\usepackage{csquotes}
\usepackage{nag}
\usepackage{flushend}
\usepackage{pifont}

\usepackage[maxnames=4,sorting=none]{biblatex}
\bibliography{literature}

\usetikzlibrary{arrows,calc,fit,shapes,arrows,shadows}

\newcommand{\pt}{\textsf{ScrambleSuit}}

\newcounter{notesctr}
\numberwithin{enumi}{section}

\numberwithin{notesctr}{section}

\hypersetup{
	colorlinks=true,
	urlcolor=blue,
	linkcolor=blue,
	citecolor=blue
}

\title{
	ScrambleSuit: A Polymorph Network \\
	Protocol to Circumvent Censorship
}

\begin{document}

\numberofauthors{3}

\author{
	\alignauthor
		Philipp Winter \\[0.2cm]
		\affaddr{Karlstad University}
	\alignauthor
		Tobias Pulls \\[0.2cm]
		\affaddr{Karlstad University}
	\alignauthor
		Juergen Fuss \\[0.2cm]
		\affaddr{Upper Austria University of \\Applied Sciences}
}

%\author{Philipp Winter\inst{1} \and Juergen Fuss\inst{2}}
%\author{Philipp Winter \and Juergen Fuss}

%\institute{
%	Karlstad University  \and Upper Austria University of Applied Sciences
%}

\maketitle

\begin{abstract}
% motivation: lots of dpi and sophisticated attacks
Deep packet inspection technology became a cornerstone of Internet censorship by facilitating cheap
and effective filtering of what censors consider undesired information. Moreover, filtering is not
limited to simple pattern matching but makes use of sophisticated techniques such as active probing
and protocol classification to block access to popular circumvention tools such as Tor.

% what we do: propose protocol which can defend against these and more attacks
In this paper, we propose \pt{}; a thin protocol layer above TCP whose purpose is to obfuscate the
transported application data. By using morphing techniques and a secret exchanged out-of-band, we
show that \pt{} can defend against active probing and other fingerprinting techniques such as
protocol classification and regular expressions.

% result: good obfuscation and little overhead
We finally demonstrate that our prototype exhibits little overhead and enables effective and
lightweight obfuscation for application layer protocols.
\end{abstract}

\keywords{Tor, bridge, pluggable transport, active probing, censorship, circumvention}

% % % % % % % % % % % % % % % % % % % % % % % % % % % % % % % % % % % % % % % % % % % % % % % % % %
\section{Introduction}
\label{sec:introduction}
% % % % % % % % % % % % % % % % % % % % % % % % % % % % % % % % % % % % % % % % % % % % % % % % % %

% Motivation: DPI is harmful -> that's why we worked on this.
We consider deep packet inspection (DPI) harmful. While originally meant to detect attack signatures
in packet payload, it is ineffective in practice due to the ease of evasion
\cite{Ptacek1998,Niemi2012,Handley2001}. At the same time, DPI technology is increasingly used by
censoring countries to filter the free flow of information or violate network neutrality
\cite{Dischinger2008}. We argue that what makes DPI particularly harmful is the \emph{asymmetry of
blocking effectiveness}, i.e., it is hard to stop motivated and skilled network intruders but very
easy to censor ordinary user's Internet access. DPI technology ultimately fails to protect critical
targets but succeeds in filtering the information flow of entire countries.

% Examples of where DPI is used.
Numerous well-documented cases illustrate how DPI technology is used by censoring countries. Amongst
others, China is using it to filter HTTP \cite{Clayton2006} and rewrite DNS responses
\cite{Anonymous2012}. Iran is known to use DPI technology to conduct surveillance \cite{irandpi}. In
Syria, DPI technology is used for the same purpose \cite{effdpi}. Even more worrying, SSL
interception proxies, an increasingly common feature of DPI boxes, are used to transparently decrypt
and inspect SSL sessions which effectively breaks SSL's confidentiality and given the rise of opaque
Internet traffic \cite{White2013}, there is no reason to believe that this trend will decrease.

% What is happening to counter this trend?
The rise of Internet censorship led to the creation of numerous circumvention tools which engage in
a rapidly developing arms race with the maintainers of censorship systems. Of particular interest to
censoring countries is the Tor network \cite{Dingledine2004}. While originally designed as a
low-latency anonymity network, it turned out to be an effective tool to circumvent censorship. The
growing success of Tor as circumvention tool did not remain unnoticed, though. Tor is or was
documented to be blocked in many countries including Iran \cite{Iran}, China \cite{Winter2012} and
Ethiopia \cite{Ethiopia}, just to name a few. We argue that many circumvention tools---Tor
included---suffer from two shortcomings which can easily be exploited by a censor.

% Active probing.
First and most importantly, they are vulnerable to \emph{active probing} as pioneered by the Great
Firewall of China (GFW) \cite{Winter2012}: the GFW is able to block Tor by first identifying
potential Tor connections based on the TLS client cipher list. If such a signature is found on the
wire, the GFW reconnects to the suspected Tor bridge and tries to ``speak'' the Tor protocol with
it. If this succeeds, the GFW blacklists the respective bridge. Active probing is not only used to
discover Tor but---as we will discuss---also VPN \cite{gfwvpn} and obfs2 \cite{obfs2probe}, which is
a censorship-resistant protocol. From a censor's point of view, active probing is a promising
strategy which greatly reduces collateral damage caused by inaccurate signatures. Also, active
probing is non-trivial to defend against because censors can easily emulate real computer users.

\begin{wrapfigure}{r}{0.15\textwidth}
	\vspace{-20pt}
	\begin{center}
		\includegraphics[width=0.15\textwidth]{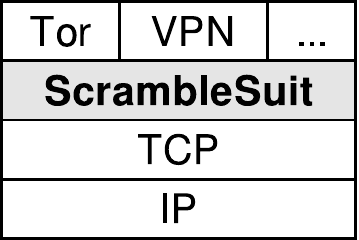}
	\end{center}
	\caption{\pt{}'s protocol stack.}
	\label{fig:protocol_stack}
	\vspace{-10pt}
\end{wrapfigure}

% Circumvention tools only have one shape.
Second, circumvention tools tend to exhibit a certain ``flow signature'' which typically remains
static. An example is Tor's characteristic 586-byte signature (cf. \S
\ref{sec:blocking_resistance}).
% TODO - start ultrasurf and look at packet sizes.
If a censor is able to deploy high-accuracy classifiers trained to recognise these very
flow signatures, the respective protocol is blocked. Censorship-resistant protocols are unable to
evade these filters by changing their flow signature.

% Circumvention tools have no way to recover from blocking -> they are not polymorph.
% Second, once a circumvention tool's protocol signature can be identified by DPI boxes, the
% respective tool becomes effectively useless. Circumvention tools typically \emph{can not recover
% from blocks}. Developers have to analyse the block, patch the software to evade the updated filter
% and finally redeploy it. This is a tedious process which can take months to ``unblock'' a tool.

% What are we doing?
In this work, we present \pt{}; a blocking-resistant transport protocol which tackles the two above
mentioned problems. \pt{} defines a thin protocol layer on top of TCP which provides lightweight
obfuscation for the transported application layer protocol. As shown in Figure
\ref{fig:protocol_stack}, \pt{} is independent of its application layer protocol and works with any
application supporting SOCKS. As a result, we envision \pt{} to be used by, among other protocols,
Tor and VPN to tackle the GFW's most recent censorship upgrades.

% Our protocol's properties.
In particular, \pt{} exhibits the following four features:
\begin{description}
	\item[Pseudo-random payload:] To an observer, \pt{}'s entire traffic is computationally
		indistinguishable from randomness. As a result, there are no predictable patterns which
		would otherwise form suitable DPI fingerprints. This renders regular expressions for the
		purpose of identifying \pt{} useless.
	\item[Polymorph:] Despite the pseudo-random traffic, a censor could still block our protocol
		based on flow characteristics such as the packet length distribution. \pt{} is, however,
		able to change its shape to make it harder for classifiers to exploit flow characteristics.
	\item[Shared secret:] We defend against active probing by making use of a secret which is shared
		between client and server and exchanged out-of-band. The server only answers to requests if
		knowledge of the secret is proven by the client.
		%The pseudo-random traffic, polymorphism and the puzzle together
		%provide undetectability as defined by Pfitzmann and Hansen \cite{Pfitzmann10}.
		%\pt{} does not hide the fact that communication happens but it hides \emph{what protocol} is
		%used for communication.
		% TODO - is this true?
	\item[Usable:] We seek to maximise \pt{}'s usability. Our protocol easily integrates in Tor's
		existing ecosystem and does not require architectural changes. Furthermore, the moderate
		protocol overhead, as shown in \S \ref{sec:experimental_evaluation}, facilitates
		comfortable web surfing.
\end{description}

% Why do we not mimic other protocols?
Blocking-resistant protocols can be split into two groups. While the first group strives to
\emph{mimic typically whitelisted protocols} such as HTTP \cite{Weinberg2012} and Skype
\cite{Moghaddam2012}, the second group aims to \emph{look like randomness}
\cite{obfs2,obfs3,Wiley2011}. Randomised protocols have the shortcoming of not being able to survive
a whitelisting censor. Nevertheless, we decided in favour of randomising because mimicing comes at
the cost of high overhead, is difficult to do correctly \cite{Houmansadr2013b} and we consider
whitelisting on a nation scale---at least for most countries---unlikely even though it is often
done in corporate networks. So, instead of maximising obfuscation while maintaining an acceptable
level of usability, we \emph{maximise usability} while keeping an \emph{acceptable level of
obfuscation}.

% Our contributions.
The contributions of this paper are as follows.
\begin{itemize}
	\item We propose \pt{}, a blocking-resistant transport protocol.
	\item We propose two authentication mechanisms based on shared secrets and polymorphism as a
		practical defence against active probing and protocol classifiers.
	\item We implement and evaluate a fully functional prototype of our protocol.
\end{itemize}

% There are no unblockable protocols.
We finally point out that unblockable network protocols do not exist. After all, censors could
always ``pull the plug'' as it was already done in Egypt \cite{Dainotti2011} and Syria
\cite{SyriaOffline}. By proposing \pt{}, we do not claim to end the arms race in our favour but
rather to raise the bar once again.

% Structure of the paper.
The remainder of this paper is structured as follows. In \S \ref{sec:related_work} we discuss
related work which is then followed by an architectural overview in \S
\ref{sec:architectural_overview}. \S \ref{sec:protocol_design} then discusses \pt{}'s design in
detail. The protocol is then evaluated in \S \ref{sec:experimental_evaluation} and the results
discussed in \S \ref{sec:discussion}. We finally conclude the paper in \S \ref{sec:conclusion}.

% % % % % % % % % % % % % % % % % % % % % % % % % % % % % % % % % % % % % % % % % % % % % % % % % %
\section{Related Work}
\label{sec:related_work}
% % % % % % % % % % % % % % % % % % % % % % % % % % % % % % % % % % % % % % % % % % % % % % % % % %
\subsection{Protocol Identification} The identification of protocols is typically motivated by quality
of service, traffic shaping and accounting -- but also censorship. In order to block protocols, they
have to be identified first. Many protocol identification techniques fail in the face of protocols
which make an active effort to remain undetected. This led to the research community finding ways
to, e.g., detect protocol tunneling in HTTP and SSH \cite{Dusi2009}, the Skype protocol
\cite{Bonfiglio2007} or encrypted traffic \cite{BarYanai2010}.

% spid + breaking & improving prot. obfuscation
Of particular relevance is the work of Hjelmvik and John \cite{Hjelmvik2010}. The authors
investigated to which extent supposedly obfuscated protocols such as Skype, BitTorrent's message
stream encryption and Spotify can be identified. Based on their findings, Hjelmvik and John suggest
evasion techniques for protocol designers which should make it harder to identify obfuscated
protocols. Some of our design decisions were motivated by their suggestions.
% wiley's evaluation framework
Similar to Hjelmvik and John, Wiley proposed a framework to dynamically classify network protocols
based on Bayesian models \cite{Wiley2011b}. This is an important first step towards the ability to
compare and evaluate blocking-resistant transport protocols.

\subsection{Protocol Obfuscation} The Tor project developed a blocking-resistant protocol called obfs2
\cite{obfs2}. The protocol implements an obfuscation layer on top of TCP and transports Tor traffic.
A passive man-in-the-middle (MITM), however, can decrypt obfs2 traffic. The successor, obfs3
\cite{obfs3}, uses a customised Diffie-Hellman handshake to solve this problem. However, both, obfs2
and obfs3 can be actively probed and do not disguise flow properties. In fact, the GFW is already
blocking obfs2 bridges by actively probing them \cite{obfs2probe}. Later in this paper, we extend
obfs3's handshake to be resistant against active probing.

% dust
Wiley's Dust protocol \cite{Wiley2011} compares to obfs2 and obfs3 in that Dust payload looks like
random data. The key exchange is handled out-of-band.  Dust also employs packet padding to
camouflage packet lengths. However, unlike \pt{}, Dust does not consider inter arrival times.
% TODO - there must be more in what we are doing!

% stegotorus
Weinberg et al. presented StegoTorus \cite{Weinberg2012}, a framework for obfuscation modules
similar to obfsproxy which is developed by the Tor project \cite{obfsproxy}. StegoTorus can
complicate protocol identification on the application layer as well as on the transport layer. Tor
connections can be multiplexed over multiple TCP connections and the application layer is
camouflaged by mimicing a cover protocol such as HTTP.

% skypemorph
SkypeMorph, as presented by Moghaddam et al. \cite{Moghaddam2012} compares to StegoTorus in that it
disguises Tor traffic by mimicing an existing protocol; in this particular case Skype video traffic.
As long as the censor does not decide to block the cover protocols, SkypeMorph and StegoTorus are
able to survive a whitelisting censor. \pt{} differs from SkypeMorph and StegoTorus since it does
not mimic a cover protocol. In fact, Houmansadr et al. claim  that protocol mimicing---as opposed to
tunneling---is a flawed approach due to the immense difficulty of mimicing a protocol correctly
\cite{Houmansadr2013b}. The authors showed that SkypeMorph and StegoTorus differ from their
respective cover protocols in numerous ways.

% FTE
Many blocking-resistant tools blindly employ different obfuscation strategies in the hope to stay
under the radar. Dyer et al. suggest the opposite \cite{Dyer2012}. The authors propose to
actively learn the regular expressions used by DPI boxes. This knowledge is then used to map cipher
text to regular expressions which are guaranteed to pass the filters. This requires the regular
expressions of DPI boxes to be known which is typically difficult in practice.

% DEFIANCE
Lincoln et al. proposed DEFIANCE \cite{Lincoln2012}: an architecture to protect Tor bridges from
being probed and their respective descriptors\footnote{A bridge descriptor is essentially a tuple
containing the bridge's IP address, port and fingerprint.} from being harvested by crawlers. The
authors accomplish these goals by developing a novel rendezvous protocol as well as a technique
called address-change signaling.

% should we also cover end point blocking here?
A solution to the problem of IP address blocking is provided by Fifield et al. \cite{Fifield2012}.
Instead of relying on long-lived static bridge IP addresses, the authors propose to use short-lived
proxies which are run by web users visiting special cooperating web sites. A practical problem
remains to be solved, however: clients making use of these so-called flash proxies must be able to
accept incoming TCP connections. This is not always possible with censored users behind NAT boxes.

\subsection{Undetectable Authentication}
Vasserman et al. proposed an undetectable authentication system based on port knocking
\cite{Vasserman2007}. Their system, SilentKnock, does however have operating system dependencies and
does not protect against connection hijacking.

Smits et al. adapted SilentKnock to better work with Tor bridges \cite{Smits2011}. The result is
called BridgeSPA. When using BridgeSPA, clients can authenticate themselves towards a bridge with
just a TCP SYN segment. If the authentication does not succeed, the bridge does not respond with a
SYN/ACK segment and the bridge appears to be offline. Just like SilentKnock, BridgeSPA does not
protect against connection hijacking and faces a number of practical problems such as the inability
to cope with NAT and the dependence on Linux kernels. While \pt{} can not hide its ``aliveness'', it
is not hindered by NAT or connection hijacking.

\begin{figure}[t]
\centering
\includegraphics[width=.35\textwidth]{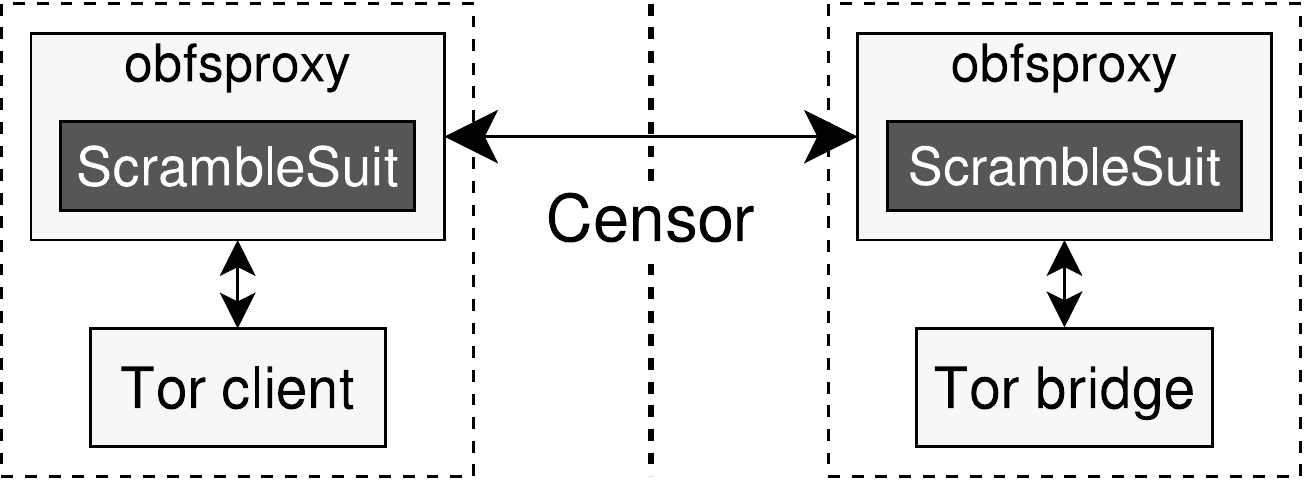}
\caption{\pt{} is a module for obfsproxy which provides a SOCKS interface for local applications.
The traffic between two obfsproxy instances is disguised by \pt{}.}
\label{fig:big_picture}
\end{figure}

\begin{figure}[t]
\centering
\includegraphics[width=.35\textwidth]{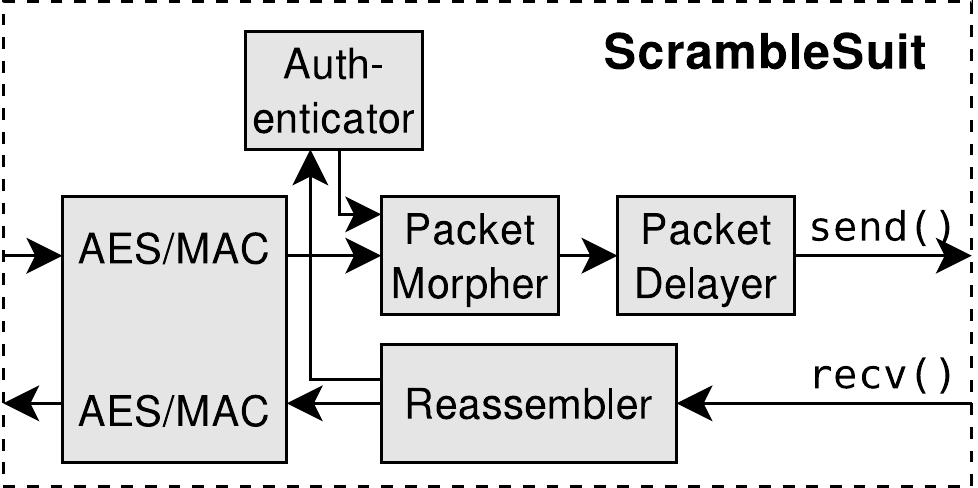}
\caption{Internally, \pt{} handles authenticated encryption of application data, client
authentication as well as flow reshaping using a packet morpher and delayer.}
\label{fig:small_picture}
\end{figure}

% % % % % % % % % % % % % % % % % % % % % % % % % % % % % % % % % % % % % % % % % % % % % % % % % %
\section{Architectural Overview}
\label{sec:architectural_overview}
% % % % % % % % % % % % % % % % % % % % % % % % % % % % % % % % % % % % % % % % % % % % % % % % % %
\pt{} is a module for obfsproxy which is an obfuscation framework developed by the Tor
project \cite{obfsproxy}. As long as obfsproxy is running on the censored client as well as on the
server, network traffic in between both communication points can be obfuscated as dictated by the
obfuscation modules. As illustrated in Figure \ref{fig:big_picture}, obfsproxy acts as a proxy
between the Tor client and the Tor bridge. While specifically designed for Tor, obfsproxy can be
used by any application as long as it supports the SOCKS protocol.

% Internal architecture.
Internally, \pt{} is composed of several components which are depicted in Figure
\ref{fig:small_picture}. Outgoing network data is first encrypted and then chopped into pieces (or
padded) by the packet morpher. Before these pieces are then sent over the wire, the packet delayer
uses small sleep calls to disguise inter arrival times. Finally, incoming network data is first
reassembled to complete \pt{} protocol messages and after decryption finally passed on to the local
application.

% What we are trying to disguise.
We aim to conceal several aspects of Tor's communication characteristics. We chose these
characteristics based on the work by Hjelmvik and John \cite{Hjelmvik2010,Hjelmvik2009} as well as
current DPI capabilities.
\begin{description}
	\item[Payload] By encrypting all \pt{} traffic, we eliminate all payload fingerprints such
		as Tor's TLS cipher list \cite{Winter2012}.
	\item[Packet length distribution] Among other things, we seek to get rid of Tor's characteristic
		586-byte packets \cite{Weinberg2012,Kadianakis2012}. We do so by morphing Tor's packet
		length distribution to a randomly chosen distribution.
	\item[Inter arrival times] Similar to the packet length obfuscation, we camouflage the inter
		arrival times by employing small and random sleep intervals before writing data on the wire.
% TODO - not possible right now
% 	\item[Packet directions] We randomise the packet directions of the first handful of packets being
% 		sent back and forth between the client and the server.
\end{description}

\subsection{Threat Model}
% Who is the adversary, what does he try to prevent and how is it blocked?
Our adversary is a \emph{nation-state censor} who desires to block unwanted network protocols and
services which would otherwise allow users within the censoring regime the retrieval of unfiltered
information or to evade the national filtering system. The censor is making use of payload analysis,
flow analysis as well as active probing to identify and then block undesired protocols.

% the censor's capabilities
The censor further has full \emph{active and passive} control over the national network. The censor
can passively monitor all traffic entering and leaving its networks in line rate. We further expect
the censor to actively tamper with traffic; namely to inject, drop and modify traffic as well as
hijack TCP sessions. We further expect the censor to select a subset of suspicious traffic for
further inspection on the slow path\footnote{We define the \emph{slow path} as the minute analysis
of a small traffic subset as opposed to the \emph{fast path} which covers the majority of all
network traffic and, as a result, has to be processed quickly.}. This could involve \emph{active
probing} as done by the GFW in order to block the Tor network \cite{Winter2012}. We model our censor
to also conduct active MITM attacks. While we believe that passive analysis and active probing
are significantly easier to deploy, there is evidence that censors are starting to---or at least
have the ability to---conduct active MITM attacks as well~\cite{gfwmitm}.

% censor can use classifiers
Our adversary is also training and deploying \emph{statistical classifiers} to identify and block
protocols. While computationally expensive, it would be imaginable that a censor uses this strategy
at least on the slow path and perhaps even on the fast path when using inexpensive flow features.

\subsubsection{Adversary Limitations}
% the censor is not whitelisting due to economical constraints
We expect the censor to be subject to economical constraints. In particular, we assume that the
censor is not using a whitelisting approach meaning that only well-defined protocols pass the
national filter. Whitelisting implies significant \emph{over-blocking} and we expect this approach
to collide with the censor's economical incentives. We also expect the censor to not block protocols
when there is only weak evidence for the protocol being blacklisted. This is a direct consequence of
avoiding over-blocking by minimising collateral damage.

% no control over users computers
Finally, we assume that the censor does not have access to or can otherwise influence censored
users' computers. Once again, we believe that such a scenario is likely to occur in corporate
networks but not on a national scale.

% % % % % % % % % % % % % % % % % % % % % % % % % % % % % % % % % % % % % % % % % % % % % % % % % %
\section{Protocol Design}
\label{sec:protocol_design}
% % % % % % % % % % % % % % % % % % % % % % % % % % % % % % % % % % % % % % % % % % % % % % % % % %
This section will discuss \pt{}'s defence against active probing, its encryption, encoding and
header format as well as how we achieve polymorphism.

% % % % % % % % % % % % % % % % % % % % % % % % % % % % % % % % % % % % % % % % % % % % % % % % % %
\subsection{Thwarting Active Probing}
\label{sec:thwarting}
% What we do in a nutshell.
We defend against active probing by proposing two mutual \emph{authentication mechanisms} which rely
on a secret which is shared \emph{out-of-band}. A \pt{} connection can only be established when both
parties can prove knowledge of this very secret. While our first authentication mechanism (see
\S \ref{sec:tickets}) is designed to work well in Tor's ecosystem, our second mechanism (see
\S \ref{sec:uniformdh}) provides additional security and efficiency if \pt{} is used by
other application protocols such as VPN.

% We already have an out-of-band channel.
With respect to Tor, there \emph{already exists} an out-of-band communication channel which is used
to distribute bridge descriptors to censored users. Naturally, we make use of this channel. If,
however, \pt{} is used to tunnel protocols other than Tor, users have to handle out-of-band
communication themselves.

% % % % % % % % % % % % % % % % % % % % % % % % % % % % % % % % % % % % % % % % % % % % % % % % % %
\subsubsection{Proof-of-Work (Again) Proves Not to Work}
\label{sec:proof-of-work}
% Why client puzzles look great.
Before deciding in favour of using a secret exchanged out-of-band, we investigated the
suitability of client puzzles. Puzzles---a variant of proof-of-work schemes---could be used by a
server to time-lock a secret. This secret can then only be unlocked by clients by spending a
moderate amount of computational resources on the problem. One particular puzzle construction,
namely time-lock puzzles as proposed by Rivest et al. \cite{Rivest1996}, provides appealing
properties such as deterministic unlocking time, asymmetric work load and inherently sequential
computation which means that adversaries in the possession of highly parallel architectures have no
significant advantage over a client with a single CPU.

% ...and why they aren't great.
While \emph{a single} client puzzle can not be solved in parallel, a censor is able to solve
\emph{multiple} puzzles in parallel by assigning all puzzles to the available CPU cores. This is
problematic because our threat model includes censors with powerful and parallel architectures.
After estimating the Tor bridge churn rate, we came to the conclusion that client puzzles would
probably not be able to increase a well-equipped censor's work load beyond the point of becoming
\emph{impractical}; at least not without becoming impractical for \emph{clients as well}. This
balancing problem is analogous to why proof-of-work schemes are believed to be unpractical for the
spam problem as well \cite{Laurie2004}.

% Why this is a better tradeoff the computational puzzles.
In summary, proof-of-work schemes would not require a shared secret but we believe that this small
usability improvement would come at the cost of greatly reduced censorship resistance. A censor in
the possession of powerful computational resources would certainly be slowed down but could
ultimately not be stopped. Active probing would simply become a matter of investing more resources.

% % % % % % % % % % % % % % % % % % % % % % % % % % % % % % % % % % % % % % % % % % % % % % % % % %
\subsubsection{Session Tickets}
\label{sec:tickets}
We now discuss the first of our two authentication mechanisms. A client can authenticate herself
towards a \pt{} server by redeeming a \emph{session ticket}. A session ticket needs to be obtained
only once out-of-band. Subsequent connections are then bootstrapped using tickets issued by the
server during the respective previous connection. A real world analogy would be a person redeeming a
ticket in order to gain access to a football stadium. Upon entering the stadium (i.e., successful
authentication), the guards give the person a new ticket so that she is able to return for the next
match. The same procedure then happens for the next match.

Session tickets are standardised in RFC 5077~\cite{rfc5077} and part of TLS since version 1.0. We
employ only a subset of the standard since we do not need its full functionality.

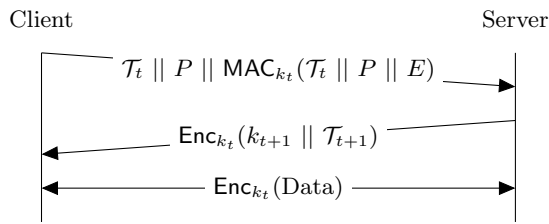
\begin{figure}
\centering
	\begin{tikzpicture}[scale=0.9]
		\draw (0,0) -- (0,2.5);
		\draw (7,0) -- (7,2.5);

		\draw [arrows={-triangle 45}]
		(0, 2.5) -- (7, 2) node [midway, fill=white, text centered]
		{ $\mathcal{T}_{t}\ ||\ P\ ||\ \textsf{MAC}_{k_{t}}(\mathcal{T}_{t}\ ||\ P\ ||\ E)$ };

		\draw [arrows={triangle 45-}]
		(0, 1) -- (7, 1.5) node [midway, fill=white, text centered]
		{ $\textsf{Enc}_{k_{t}}(k_{t+1}\ ||\ \mathcal{T}_{t+1})$ };

		\draw [arrows={triangle 45-triangle 45}]
		(0, 0.5) -- (7, 0.5) node [midway, fill=white, text centered]
		{ $\textsf{Enc}_{k_{t}}$(Data) };

		\node at (0,3) {Client};
		\node at (7,3) {Server};
	\end{tikzpicture}
	\caption{The client redeems a valid session ticket $\mathcal{T}_{t}$ containing the master key
	$k_{t}$. The server responds by issuing a new ticket $\mathcal{T}_{t+1}$ for future use. Both
parties then exchange application data.}
\label{fig:tickets}
\end{figure}

% The basic idea.
The basic idea is illustrated in Figure \ref{fig:tickets}. \pt{} servers issue new session tickets
$\mathcal{T}_{t+1}$ which contain a future shared master key $k_{t+1}$ and an issue date $d$ indicating
the ticket's creation time. Session tickets are encrypted and authenticated with secret keys
$k_{S}$\footnote{For simplicity, we refer to these two symmetric keys as just $k_{S}$ while they are
in fact two keys: one for encryption and one for authentication.} only known to the server, i.e.,
$\mathcal{T}_{t+1} = \textsf{Enc}_{k_{S}}(k_{t+1}\ ||\ d)$. As a result, a ticket $\mathcal{T}$ is
opaque to the client. Note that a client, when obtaining a ticket, also has to learn the master key
$k_{t+1}$ in order to be able to derive the same session keys as the server; so clients always
obtain the tuple ($k_{t+1}\ ||\ \mathcal{T}_{t+1})$. Session tickets have the advantage that the
server \emph{does not have to keep track} of issued tickets. Instead, the server's state is
outsourced and stored by clients which greatly reduces a server's load.

% New tickets are issued upon connection.
Whenever a client successfully connects to a \pt{} server, the server issues a new ticket
concatenated to the according master key ($k_{t+1}\ ||\ \mathcal{T}_{t+1}$) for the client. The
tuple is placed in a special \pt{} control message (see \S \ref{sec:confidentiality}). The new
ticket is sent immediately after successful bootstrapping.

% How long are tickets valid?
We mentioned earlier that a \pt{} server manages secret keys $k_{S}$ which are used to encrypt and
authenticate session tickets. This prevents clients from tampering with tickets and the server can
verify that a newly received and authenticated ticket was, in fact, issued by the server. Servers
rotate their $k_{S}$ keys after a period of seven days. After the generation of new $k_{S}$ keys,
the superseded keys are kept for another seven days in order to decrypt and verify (but not to
issue!) tickets which were issued by the superseded keys. As a result, tickets are always
valid and redeemable for a period of \emph{exactly seven days}; no matter when they were issued. As
a result, as long as a user keeps reconnecting to a \pt{} server at least once a week, \emph{key
continuity} is ensured and there is no need for additional out-of-band communication.

% Length obfuscation.
A censor could now conduct traffic analysis by looking for TCP connections which always begin with
the client sending $|\mathcal{T}|$ bytes to the server. To obfuscate the ticket length, we introduce
random padding $P$ and authenticate the ticket $\mathcal{T}$ as well as the padding $P$ by computing
$\textsf{MAC}_{k_{t}}(\mathcal{T}\ ||\ P\ ||\ E)$ with $k_{t}$ being the shared master secret which
the client obtained together with the ticket and $E$ discussed in the following paragraph. Both
parties will use $k_{t}$ to derive session keys as discussed in \S \ref{sec:confidentiality}. The
server knows that all bytes of the handshake were successfully received when the last bytes form a
valid MAC over the previous bytes. The exact amount of random padding is determined by the packet
morpher discussed in \S \ref{sec:packet_lengths}. We use HMAC-SHA256-128 for the MAC.

% How we prevent replay attacks.
At this point, a censor could still intercept tickets and replay them. This would make the server
issue a new ticket for the censor. While the censor would not be able to read the resulting \pt{}
control message---the shared master key $k_{t}$ would be unknown---the fact that a replay attack
triggers a response can be suspicious. We prevent replay attacks, or in other words ticket double
spending, by caching master keys $k_{t}$. If a server encounters a cached $k_{t}$, it does not reply
to prevent the censor from learning the server's state. We begin to cache a $k_{t}$ after a new session
ticket was issued and the client acknowledged that she correctly received the new ticket by using a
special \pt{} message type (see \S \ref{sec:confidentiality}). To reduce the amount of keys to
cache, we add the value $E$ to the MAC. It refers to the Unix epoch divided by 3600, i.e., the
current time with a granularity of one hour. While this requires client and server to have loosely
synchronised clocks, the server has to cache redeemed keys for a period of only one hour instead of
seven days.

Session tickets already provide a strong level of protection. Active probing and replay attacks are
foiled while forward secrecy is provided. Therefore, we envision session tickets to be satisfactory
for most application protocols to be tunneled over \pt{}.

\begin{figure}
\centering
\includegraphics[width=0.47\textwidth]{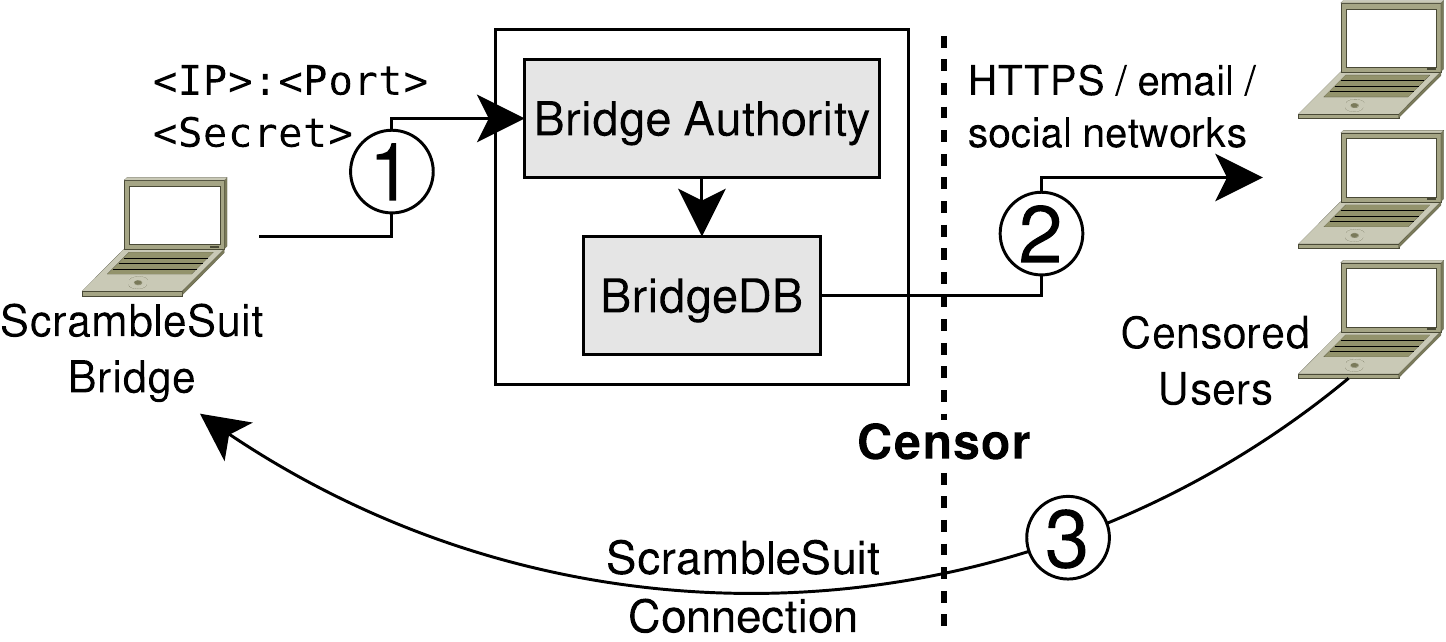}
\caption{\pt{} bridges send their descriptor to the bridge authority \ding{192}. From there, it is
distributed to censored users who learn about IP address, port and the secret \emph{out-of-band}
\ding{193}. Finally, direct connections can be established \ding{194}.}
\label{fig:scramblesuit-tor}
\end{figure}

% Why session tickets are not sufficient for the public Tor distribution channels.
Session tickets do not, however, integrate well with Tor's existing ecosystem. The reason lies in
how Tor bridges are distributed to users. The process is illustrated in Figure
\ref{fig:scramblesuit-tor}. Volunteers will set up \pt{} bridges which then publish their
descriptors---including IP address, port and secret---to the bridge authority (1) which feeds this
information into the BridgeDB component. In the next step, the gathered descriptors have to be
distributed to censored users (2). The two primary distribution channels are email and HTTPS
\cite{Ling2012}.  Users can ask for bridges over email or they can visit the bridge distribution
website\footnote{URL: \url{https://bridges.torproject.org}.} and obtain a set of bridges after
solving a CAPTCHA. The problem is that \emph{one bridge descriptor} is typically shared by
\emph{many users}.  All these users would end up with an identical session ticket. This causes two
severe problems.  First, our replay protection mechanism does not allow reuse of session tickets.
Second, session ticket reuse would lead to identical byte strings at the beginning of a \pt{}
handshake which would be a strong distinguisher. These problems lead us to our \emph{second
authentication mechanism} which is optimised for Tor and can function with a secret which is shared
by many users as shown in the scenario in Figure \ref{fig:scramblesuit-tor}.

% % % % % % % % % % % % % % % % % % % % % % % % % % % % % % % % % % % % % % % % % % % % % % % % % %
\subsubsection{Uniform Diffie-Hellman}
\label{sec:uniformdh}
Our second authentication mechanism is an extension of the Uniform Diffie-Hellman (UniformDH)
handshake which was proposed in the obfs3 protocol specification \cite[\S 3]{obfs3}. obfs3's
handshake makes use of uniformly distributed public keys which are only negligibly different from
random bytes. As a result, UniformDH can be used to agree on a master key $k_{t}$ without a censor
knowing that Diffie-Hellman is used.

% How UniformDH works.
UniformDH is based on the 1536-bit modular exponential group defined in RFC 3526 \cite{rfc3526}.
When initiating a UniformDH handshake, the client first generates a 1536-bit private key $x$. The
least significant bit of $x$ is then unset in order to make the number even. The public key $X$ is
defined as $X = g^{x}$ (mod $p$) where $g = 2$. The server computes its private key $y$ and its
public key $Y$ the same way. To prevent a censor from learning that $X$ is a quadratic residue mod
$p$---a clear distinguisher---the client randomly chooses to send either $X$ or $p - X$ to the
server. The server can then derive the shared master secret by calculating $k_{t} = X^y$ (mod $p$).
Since the private keys $x$ and $y$ are even, the exponentiations $X^y$ (mod $p$) and $(p - X)^y$
(mod $p$) result in the same shared master secret.

% % % % % % % % % % % % % % % % % % % % % % % % % % % % % % % % % % % % % % % % % % % % % % % % % %
\subsubsection{Extending Uniform Diffie-Hellman}
% Why we need to extend UniformDH.
In its original form, the UniformDH construction does not protect against active probing. A censor
who suspects UniformDH can simply probe the supposable bridge and opportunistically initiate a
UniformDH handshake. To prevent that attack, we now turn UniformDH's anonymous handshake into an
authenticated handshake in order to be resistant against active attacks.

\begin{figure}[t]
\centering
	\begin{tikzpicture}[scale=0.9]
		\draw (0,0) -- (0,3);
		\draw (7,0) -- (7,3);

		\draw [arrows={-triangle 45}]
		(0, 3) -- (7, 2.5) node [midway, fill=white, text centered]
		{ $X\ ||\ P\ ||\ \textsf{MAC}_{k_{B}}(X\ ||\ P\ ||\ E)$ };

		\draw [arrows={triangle 45-}]
		(0, 1.5) -- (7, 2) node [midway, fill=white, text centered]
		{ $Y\ ||\ P\ ||\ \textsf{MAC}_{k_{B}}(Y\ ||\ P\ ||\ E)$ };

		\draw [arrows={-triangle 45}]
		(7, 1.5) -- (0, 1) node [midway, fill=white, text centered]
		{ $\textsf{Enc}_{k_{t}}(k_{t+1}\ ||\ \mathcal{T}_{t+1})$ };

		\draw [arrows={triangle 45-triangle 45}]
		(0, 0.5) -- (7, 0.5) node [midway, fill=white, text centered]
		{ $\textsf{Enc}_{k_{t}}$(Data) };

		\node at (0,3.5) {Client};
		\node at (7,3.5) {Server};
	\end{tikzpicture}
	\caption{After client and server agreed on the master key $k_{t}$ using Diffie-Hellman, the
	server is issuing a new session ticket for the client. Afterwards, both parties exchange
	application data.}
\label{fig:uniformdh}
\end{figure}
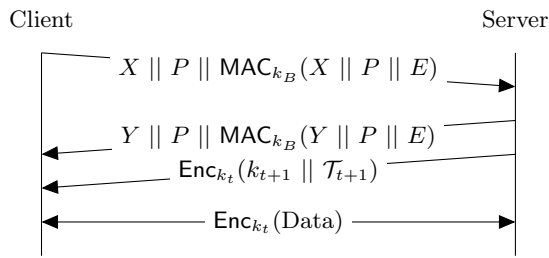

% How we extend UniformDH.
We do so quite similar to the session tickets discussed in \S \ref{sec:tickets}. As depicted in
Figure \ref{fig:uniformdh}, we concatenate pseudo-random padding $P$ and a MAC to the public keys
$X$ and $Y$.  The MAC authenticates the respective public key as well as the padding. The MAC is
keyed by a shared secret $k_{B}$ which is distributed together with the Tor bridge's IP:port
tuple over email or HTTPS (cf. step \ding{192} in Figure \ref{fig:scramblesuit-tor}). As with tickets, the
server and client know that the handshake message was fully received when the last received bytes
form a valid MAC over the previous bytes. Note that $k_{B}$ \emph{can be reused} because it is only
used to key the MAC. The handshake is conducted using UniformDH with randomly chosen public keys. As
a result, two subsequent UniformDH handshakes based on the same $k_{B}$ will appear to be different
to a censor. We defend against replay attacks by adding $E$, the Unix epoch divided by 3600, to the
MAC and cache the MAC for a period of one hour.

% Transition to tickets ASAP.
A successful UniformDH key agreement is followed by the server issuing a session ticket for the
client. The client will then redeem this ticket upon connecting to the server the next time.
Accordingly, we expect the UniformDH handshake to be done \emph{only once}, namely when a Tor client
connects to a bridge for the first time. From then one, session tickets will be used to connect to
the same bridge.

% Difference between UniformDH and tickets is negligible.
To a censor, the payload of both authentication schemes is computationally indistinguishable from
randomness. As a result, a censor who is assuming that a server is running \pt{} is unable to tell
whether a client successfully authenticated herself by using UniformDH or by redeeming a session
ticket.

% TODO - Rotate keys?

% Diffie-Hellman is weaker but better suited for Tor. Tickets should be preferred, though.
We finally stress that bootstrapping \pt{} using UniformDH provides \emph{less security} than 
when bootstrapped using session tickets. Since the secret key $k_{B}$ for UniformDH will be used by
multiple clients, a malicious client in the possession of $k_{B}$ and who is able to eavesdrop on the
connection of another client using the same \pt{} server can conduct active MITM attacks. While Tor
does protect against active MITM attacks, this can be problematic for application protocols
other than Tor. Therefore, we emphasise that session tickets are the preferred authentication
mechanism whereas our UniformDH extension's sole purpose is to make \pt{} work well in Tor's
infrastructure.

% % % % % % % % % % % % % % % % % % % % % % % % % % % % % % % % % % % % % % % % % % % % % % % % % %
\subsubsection{Usability Considerations}
% After all, everybody will use copy&paste.
In order for a user to successfully connect to a \pt{} server, she needs a \emph{triple}: an IP
address, a TCP port and a secret which is either the UniformDH secret $k_{B}$ or a session ticket
tuple $(k_{t}\ ||\ \mathcal{T}_{t})$. We expect these triples to be distributed
mostly electronically; over email, instant messaging programs or online social networks. As a
result, a user can simply copy and paste the entire triple into her obfsproxy configuration file.

% Base32 for easier verbal distribution.
We do, however, also expect limited verbal distribution of \pt{} triples, e.g., over a telephone
line. To facilitate this, we define the encoding format of secrets and tickets to be Base32. Base32
strings consist of the letters A--Z, the numbers 2--7\footnote{The numbers 0 and 1 are omitted to
prevent confusion with the letters I and O.} as well as the padding character ``=''. Since
there is no distinction between uppercase and lowercase letters, we hope to make verbal distribution
less confusing and error-prone. After all, a \pt{} bridge descriptor would look like: \texttt{Bridge
scramblesuit 1.2.3.4:443 password=NCA6I6GZZD42BWUB}. We believe that the prefix \texttt{password=}
will find more acceptance among users than simply appending the secret.

% % % % % % % % % % % % % % % % % % % % % % % % % % % % % % % % % % % % % % % % % % % % % % % % % %
\subsection{Header Format and Confidentiality}
\label{sec:confidentiality}
\begin{figure}
\centering
\includegraphics[width=0.47\textwidth]{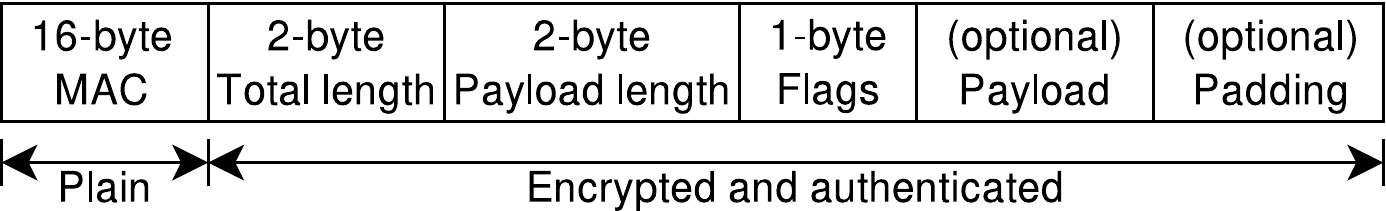}
\caption{\pt{}'s message header format. The encrypted part is authenticated by a HMAC-SHA256-128.
The entire message is computationally indistinguishable from randomness.}
\label{fig:header}
\end{figure}

% We have a custom message format.
Our protocol employs a custom message format whose header is illustrated in Figure \ref{fig:header}.
\pt{} exchanges variable-sized messages with optional padding which is discarded by the
remote machine.

% HMAC.
The first 16 bytes of the header are reserved for a HMAC-SHA256-128 which protects the integrity and
authenticity of the protocol message. In accordance with the encrypt-then-MAC paradigm, the HMAC is
computed over the encrypted remainder of the message. The secret key required by the HMAC is derived
from the shared master key.

% There are length fields.
The HMAC is followed by two bytes which specify the total length of the protocol message. \pt{}'s
maximum transmission unit is 1460-byte-sized messages. Together with a minimal IP and TCP header,
this adds up to 1500-byte packets which fill an Ethernet frame. In order to be able to distinguish
padding from payload, the next two bytes determine the payload length. If no padding is used, the
payload length equals the total length.

% Message type.
To separate application data from protocol signaling, we define a 1-byte message flag field. The
first bit signals application data in the message body whereas a message with the second bit set
contains a newly issued session ticket. The third bit (which can be set together with the first bit)
confirms the receiving of a session ticket. We reserve the remaining bits for future use. In
particular, they could be used to negotiate flow properties as discussed in \S
\ref{sec:changing_shape}.

% Payload and encryption.
The header is then followed by the message payload which contains the application protocol
transported by \pt{}. We employ encryption in order to hide the application protocol, the padding as
well as \pt{}'s header. With regard to Tor, this means that the already encrypted Tor traffic is
wrapped inside yet another layer of encryption. For encryption, we use 256-bit AES in counter mode.
The counter mode effectively turns AES into a stream cipher. We use two symmetric keys: one for the
traffic $C \to S$ and one for $S \to C$. Both symmetric keys as well as the respective nonces for
the counter mode are derived from the shared master secret using HKDF based on SHA256 \cite{hkdf}.

% % % % % % % % % % % % % % % % % % % % % % % % % % % % % % % % % % % % % % % % % % % % % % % % % %
\subsection{Changing Shape}
\label{sec:changing_shape}
% we value performance more than unblockability but change our trade-off
So far, we discussed defences against censors who analyse packet payload or conduct active attacks
to reveal \pt{}'s presence. However, a censor could also make use of \emph{traffic analysis}. In
this section, we propose lightweight countermeasures to diminish---but not to defeat!---such
attacks. In particular, we will teach \pt{} how to change its ``protocol shape''\footnote{This
happens analogous to the \emph{scramble suits} in Philip K. Dick's novel ``A Scanner Darkly''.}.

% what do we mean by 'shape'?
Our definition of \pt{}'s shape is twofold: we consider \emph{packet lengths} and \emph{inter
arrival times}. While our transported data is encrypted and exhibits no structure, these flow
metrics can still leak information about the application protocol
\cite{Crotti2007,Dyer2012b,Cai2012}. As a result, we seek to randomise these characteristics in
order to decrease the accuracy of protocol classifiers used to detect \pt{}.

% What this means for our implementation.
In general, the kernel's TCP stack is responsible for packet lengths. In order to affect packet
lengths in user space, we deactivate Nagle's algorithm which seeks to avoid unnecessarily small TCP
segments. This comes, however, at the cost of increased protocol overhead.

% how we randomly generate discrete probability distributions.
The randomisation of packet lengths as well as inter arrival times is based on a randomly generated
discrete probability distribution. We generate these distributions by first determining the amount
of bins $n$ which is uniformly chosen from the set $\{1..100\}$. In the next step, we assign each
bin $b_{i}$ for $1 \le i \le n$ a probability by randomly picking a value in the interval $]0, 1 -
\sum^{n}_{i=1} b_{i-1}[$ for $b_{0} = 0$. The following gives an example of four assignments.

\begin{flalign}\label{equ:example}
& b_{0} \gets 0 \\
& b_{1} \overset{R}{\gets}\ ]0, 1 - b_{0}[ \\
& b_{2} \overset{R}{\gets}\ ]0, 1 - b_{0} - b_{1}[ \\
& b_{n} \overset{R}{\gets}\ ]0, 1 - b_{0} - ... - b_{n-1}[
\end{flalign}
	
We will show in \S \ref{sec:experimental_evaluation} that this naive approach turns out to be
sufficient to obfuscate Tor's flow characteristics. A specialised algorithm---e.g., to optimise
throughput---would be conceivable but is beyond the scope of this paper.

% % % % % % % % % % % % % % % % % % % % % % % % % % % % % % % % % % % % % % % % % % % % % % % % % %
\subsubsection{Packet Length Adaption}
\label{sec:packet_lengths}
% Packet length leak information.
It is well known that a network flow's packet length distribution leaks information about the
network protocol \cite{Hjelmvik2010,Crotti2007,Lim2010} and even the content
\cite{Panchenko2011,Cai2012}. For instance, a large fraction of Tor's traffic contains 568-byte
packets which is the result of Tor's internal use of 512-byte cells plus TLS' header (see Figure
\ref{fig:c2spl} and \ref{fig:s2cpl}). These 568-byte packets form a strong distinguisher which can
be used to detect Tor by simply capturing a few dozen network packets as shown by Weinberg et al.
\cite{Weinberg2012}. To defend against such simple applications of traffic analysis, we modify
\pt{}'s packet length distribution.

% Why we don't use traffic morphing.
An efficient way to morph a source distribution to a target distribution was proposed
by Wright, Coull and Monrose \cite{Wright2009}. Their concept, traffic morphing, relies on the
computation of a morphing matrix to minimise the overhead when morphing a source distribution to a
target distribution. Unfortunately, we cannot make use of traffic morphing because our target
distribution is dynamic which would require frequent recomputation of the morphing matrix which is
an expensive operation. This would lead to unnecessary CPU load on the client as well as on the
bridge. Furthermore, our source distribution is not known since \pt{} is designed to be able
to handle arbitrary application protocols.

% We use naive sampling.
Instead, we adopt \emph{naive sampling} to disguise the application protocol's packet length
distribution. Every time \pt{} establishes a connection to a server, it randomly generates a fresh
discrete probability distribution as discussed earlier. Every bin in the probability distribution is
uniformly chosen from the set $\{1..1460\}$. The newly generated distribution is then randomly
sampled for every chunk of application data, \pt{} is about to send over the wire. After a sample
length is obtained, our algorithm either \emph{a)} pads the current packet to fit the sample's
length or \emph{b)} splits and sends it and then proceeds to morph the remaining data the same way.

% TODO - show how to calculate overhead.

% % % % % % % % % % % % % % % % % % % % % % % % % % % % % % % % % % % % % % % % % % % % % % % % % %
\subsubsection{Inter Arrival Time Adaption}
% Inter arrival time also leaks information.
Analogous to the packet length distribution, the distribution of inter arrival times between
consecutive packets has discriminative power and could be used by censors to identify protocols
\cite{Jaber2011}. In contrast to the packet length distribution, inter arrival times are frequently
distorted by network jitter, overloaded middle boxes and the communicating end points. Nevertheless,
we believe that it would be no sound strategy to assume the network to be unreliable enough to
render measurements useless. Therefore, \pt{} is also able to modify its inter arrival times. The
mechanism is the same as for the packet length adaption: a random distribution is generated and then
random samples are drawn from it. The samples are the parameters for short \texttt{sleep()} calls
which are invoked prior to sending data to the remote end.

% we can only increase it, so we keep it low
We are only able to increase inter arrival times but not to decrease them. Increased inter arrival
times have a direct negative effect on throughput and can easily turn into a nuisance for users when
getting too high. As a result, we keep the sleep intervals within the interval of $[0, 100[$
milliseconds. We believe that this interval represents a reasonable tradeoff between obfuscation
and throughput as we will show in \S \ref{sec:blocking_resistance}.

% % % % % % % % % % % % % % % % % % % % % % % % % % % % % % % % % % % % % % % % % % % % % % % % % %
% \subsubsection{Packet Direction Adaption}

\newpage
% % % % % % % % % % % % % % % % % % % % % % % % % % % % % % % % % % % % % % % % % % % % % % % % % %
\subsubsection{Shortcomings}
\label{sec:shortcomings}
It is important to note that for a censor armed with a well-chosen set of features, traffic analysis
can be a powerful attack and strong defences are believed to be expensive \cite{Dyer2012b,Cai2012}.
We made an effort to disguise obvious flow features while keeping throughput high enough to
facilitate comfortable web surfing and enable the transportation of low-latency applications over
\pt{}.

A censor can still measure derived flow metrics such as ``total bytes transferred'', packet
directions or the ``burstiness'' of \pt{}'s behaviour. These metrics would be expensive to disguise
and by exploiting them, a censor would at least be able to guess whether \pt{}'s transported
application is a bulk file transfer or a request-response protocol. Nevertheless, traffic analysis
does not give censors a \emph{certain answer}. False positives are always a problem and can lead to
overblocking. As mentioned in our threat model, we believe that the censor might use traffic
analysis to select a subset of traffic for closer inspection but not to block flows.

% % % % % % % % % % % % % % % % % % % % % % % % % % % % % % % % % % % % % % % % % % % % % % % % % %
\section{Experimental Evaluation}
\label{sec:experimental_evaluation}
% % % % % % % % % % % % % % % % % % % % % % % % % % % % % % % % % % % % % % % % % % % % % % % % % %

% some small facts about our prototype
We implemented a prototype of \pt{} in the form of several Python modules for obfsproxy. Our
prototype consists of approximately 2000 lines of code. The measurements below were all
conducted using this prototype.

\begin{figure}[h]
\centering
	%\vspace{-30pt}
	\begin{center}
		\includegraphics[width=0.27\textwidth]{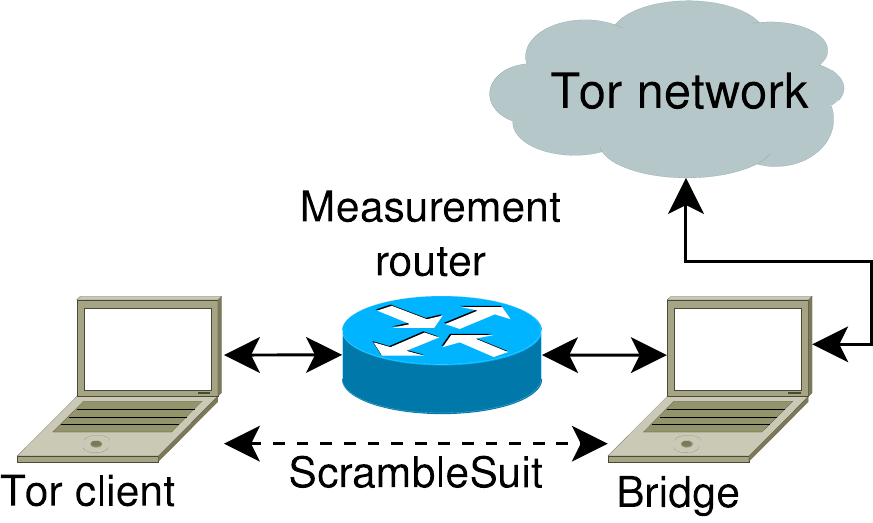}
	\end{center}
	\caption{The experimental setup.}
	\label{fig:experimental_setup}
	%\vspace{-10pt}
\end{figure}

As illustrated in Figure \ref{fig:experimental_setup}, our experimental setup consisted of two
Debian GNU/Linux machines which were connected via a router which performed the measurements. All
three machines were connected over Fast Ethernet. We expect this setup to be ideal for a censor
because it does not cause IP fragmentation or high latency due to overloaded middle boxes. As a
result, we believe that a censor would do worse in practice. Both of our machines were running Tor
v0.2.4.10-alpha and obfsproxy. The Tor bridge was configured to be private and was only used by
our client. The bridge then relayed all traffic into the public Tor network. Note that \pt{} is only
``spoken'' in between the client and the bridge.

%\begin{wrapfigure}{r}{0.27\textwidth}
%\end{wrapfigure}

% % % % % % % % % % % % % % % % % % % % % % % % % % % % % % % % % % % % % % % % % % % % % % % % % %
\subsection{Blocking Resistance}
\label{sec:blocking_resistance}
% experimental setup
To create network traffic for our measurements, we downloaded the 1 MB Linux kernel v1.0 from
kernel.org\footnote{\url{https://www.kernel.org/pub/linux/kernel/v1.0/linux-1.0.tar.bz2}.} on
the client. We downloaded the file once over Tor\footnote{In fact, we downloaded the file many times
over Tor and found that consecutive runs differed mostly in the ratio between 586-byte and 1448-byte
packets. As a result, we only plot one run.} and 5 times over \pt{}.

\begin{figure}[t]
\centering
\includegraphics[width=0.35\textwidth]{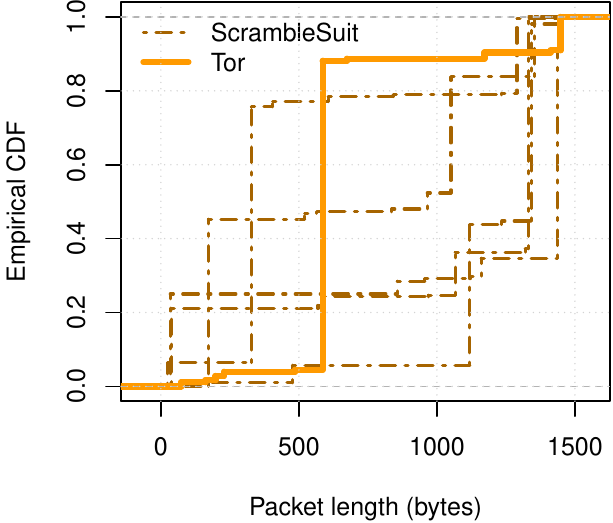}
\caption{Client-to-server packet lengths.}
\label{fig:c2spl}
\end{figure}

\begin{figure}[t]
\centering
\includegraphics[width=0.35\textwidth]{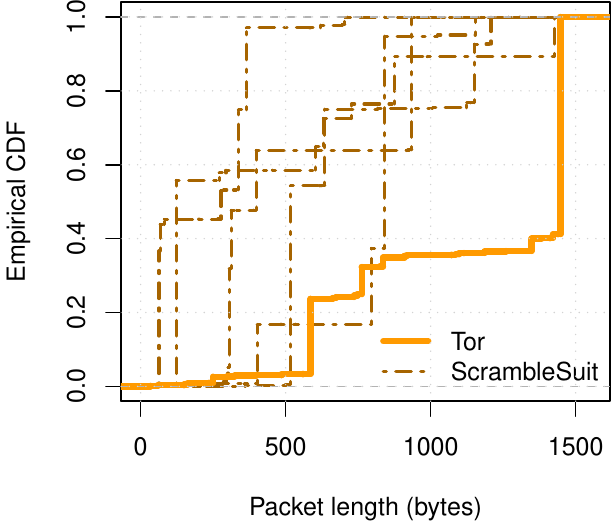}
\caption{Server-to-client packet lengths.}
\label{fig:s2cpl}
\end{figure}

% explain diagrams
The packet length distribution for client-to-server traffic is illustrated in Figure \ref{fig:c2spl}
and the server-to-client traffic in Figure \ref{fig:s2cpl}. The solid orange line represents the
download over Tor. The prevalence of 586-byte packets is clearly visible; especially for the
client-to-server traffic. These packets contain internal Tor cells which handle flow control. All
these segments had to be wrapped into a 512-byte Tor cell. In addition, more than 50\% of the
server-to-client traffic consists of 1448-byte packets. The remaining brown lines represent 5
consecutive downloads over \pt{}. Both empirical cumulative distribution functions visibly deviate
from Tor's.

\begin{figure}[t]
\centering
\includegraphics[width=0.35\textwidth]{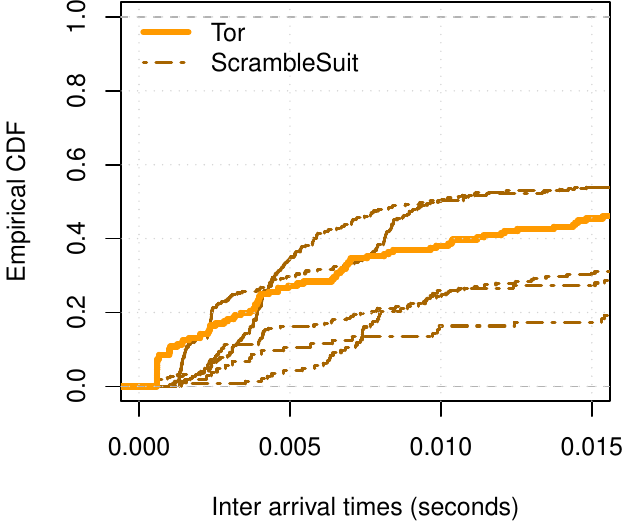}
\caption{Client-to-server inter arrival times.}
\label{fig:c2siat}
\end{figure}

\begin{figure}[t]
\centering
\includegraphics[width=0.35\textwidth]{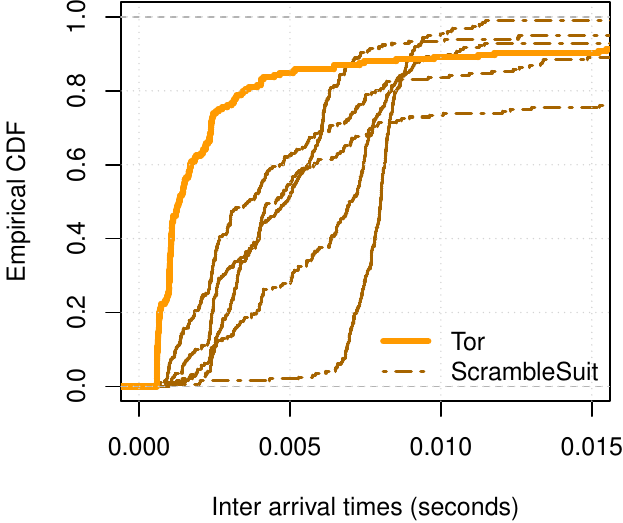}
\caption{Server-to-client inter arrival times.}
\label{fig:s2ciat}
\end{figure}

Figure \ref{fig:c2siat} and \ref{fig:s2ciat} depict the inter arrival times of the same data. The
delays in Figure \ref{fig:c2siat} tend to be rather high---only 40\% of Tor packets had an inter
arrival delay under 10 ms---because the client only acknowledged the bulk data coming from the
server. For this reason, the delays are much smaller in Figure \ref{fig:s2ciat}. For both, the
packet lengths as well as the inter arrival times, a two-sample Kolmogorov-Smirnov test rejected the
hypothesis that the Tor distributions equal any of \pt{}'s distributions.

\begin{table*}[t]
\caption{Mean ($\mu$) and standard deviation ($\sigma$) of the goodput, transferred KBytes and the
total overhead. The data was generated based on the download of a 1,000,000-byte file.}
\label{table:performance}
\begin{center}
\begin{tabular}{|c|c|c|c|c|c|c|}
\hline
& \multicolumn{2}{|c|}{\textbf{HTTP}} & \multicolumn{2}{|c|}{\textbf{Tor}} &
	\multicolumn{2}{|c|}{\textbf{\pt{}}} \\
& \multicolumn{2}{|c|}{$\mu\hspace{1.2cm}\sigma$} &
   \multicolumn{2}{|c|}{$\mu\hspace{1.2cm}\sigma$} &
   \multicolumn{2}{|c|}{$\mu\hspace{1.2cm}\sigma$} \\

\hline
\textbf{Goodput} & 6.3 MB/s & 3.4 MB/s & 279.7 KB/s & 293.9 KB/s & 89.8 KB/s & 41.1 KB/s \\
\hline
\textbf{C$\to$S KBytes} & 23.1 & 1.6 & 71.9 & 7.7 & 132.5 & 35.5 \\
\hline
\textbf{S$\to$C KBytes} & 1047 & 20.7 & 1121.6 & 38.8 & 1242.3 & 70.2 \\
\hline
\textbf{Total overhead} & 9.5\% & 2.2\% & 22.2\% & 4.4\% & 40.7\% & 10.1\% \\
\hline
\end{tabular}
\end{center}
\end{table*}

\subsection{Performance}
In order to evaluate \pt{}'s overhead and goodput, we created a 1,000,000-byte file consisting of
random bytes and placed it on a web server operated by our institution. We then downloaded the file
% TODO - s/institution/karlstad/ because mulitple institutions involved.
with \texttt{wget} 25 times over HTTP, Tor and \pt{}, respectively. For Tor and \pt{}, we
established a new circuit for every download but we used the same entry guard in order to make the
results more comparable. Based on the measured data, we calculated the mean $\mu$ and the standard
deviation $\sigma$ for several performance metrics. The results are depicted in Table
\ref{table:performance}.

% Goodput.
The goodput refers to the application layer throughput. We achieved very high values for the HTTP
download because the file transfer could be carried out over the LAN. Tor averaged at roughly 280
KB/s and \pt{} achieved approximately one third of that. Just like Tor, \pt{} exhibits a high
standard deviation. We believe that this was mostly caused by differences in circuit throughput but
\pt{} also fluctuates due to its different shapes. Shapes featuring many large packet lengths lead
to higher throughput whereas shapes with small packets tend to harm throughput.

% KBytes.
The next two rows of Table \ref{table:performance} refer to the transferred KBytes from client to
server (C$\to$S) and server to client (S$\to$C). Note that this covers all the data which was
present on the wire; including IP and TCP header. The consideration of IP and TCP overhead is
important because one of the reasons for \pt{}'s overhead is the varying packet lengths which
imply an increase in IP and TCP headers. Unsurprising, Tor transferred more data than HTTP because
of Tor's and TLS' protocol overhead. \pt{} transferred the most data because of the additional
protocol header (see \S \ref{sec:confidentiality}) as well as the varying packet lengths.
This is emphasised by the high standard deviation of 35 and 70 KBytes, respectively.

% Total overhead.
The last row illustrates the total protocol overhead. Once again, we also consider IP and TCP
headers. HTTP has the lowest overhead followed by Tor and finally \pt{}. Our protocol exhibits 40\%
overhead with a high standard deviation of 10\% which stems from the shape shifts.

% % % % % % % % % % % % % % % % % % % % % % % % % % % % % % % % % % % % % % % % % % % % % % % % % %
\section{Discussion}
\label{sec:discussion}
% % % % % % % % % % % % % % % % % % % % % % % % % % % % % % % % % % % % % % % % % % % % % % % % % %
We made an effort to design \pt{} in a way to be resistant against current censorship threats; most
notably active probing. In this section we discuss the remaining attack surface and point out
emerging problems.

\subsection{Attacks on ScrambleSuit}

\textbf{Active Probing}: A censor could still actively probe a \pt{} server. Upon
establishing a TCP connection, a censor could proceed by sending arbitrary data. However, the server
will not respond without prior authentication. In contrast to SilentKnock \cite{Vasserman2007} and
BridgeSPA \cite{Smits2011}, \pt{} does not disguise its ``aliveness''. While this approach does
leak information\footnote{A censor learns that a server is online but unwilling to talk unless given
the ``correct'' data.}, it has the benefit of making \pt{} significantly easier to deploy due to
lack of platform dependencies.

\textbf{Packet Injection, Modification and Dropping}: A censor could tamper with an existing \pt{}
connection by injecting data, modifying bytes on the wire or dropping packets. Both communicating
parties will detect injected or modified data due to the MAC being invalid. Hijacking a \pt{} TCP
connection boils down to the same problem; a censor would bypass the authentication but is unable to
talk to the \pt{} server because the session keys are unknown. Finally, dropped packets are handled
by the application protocol and could trigger TCP retransmissions.

\textbf{Payload \& Flow Analysis}: Payload analysis would only yield data which is computationally
indistinguishable from randomness. Flow analysis, on the other hand, would yield a certain
distribution of packet lengths and inter arrival times which changes from connection to connection.
While \S \ref{sec:shortcomings} showed that defending against traffic analysis can be costly, our
main objective is to thwart active probing attacks because they enable \emph{deterministic protocol
identification}. Sophisticated traffic analysis attacks will always have a range of
\emph{uncertainty}. We believe that the GFW's very reason to conduct active probing is to obtain
certainty and avoid collateral damage. Protocol blocking based on traffic analysis will unavoidably
imply false positives.

\subsection{Future Work}
% We need better bridge distribution.
Improving Tor's censorship resistance is a two-sided problem. On the one hand, bridge descriptors
need to be distributed to users while not falling into the hands of censors and on the other hand,
the subsequent Tor connection should be hard to identify for censors. While we focused on the
latter, the former remains an open problem as well. Recent work focused on reputation-based models to
maximise bridge aliveness \cite{McCoy2011,Wang2013}.

% Whitelisting.
The arms race with circumvention tools might pressure censors into introducing whitelisting. While
we are not aware of country-wide whitelists, Russia is experimenting with a ``Clean Internet''
\cite{russiawl}. Should this approach turn out to be successful for censors, the arms race will
shift towards tunneling circumvention traffic through whitelisted protocols.

% There is no ``fingerprintless''.
We finally point out that no protocol is ``fingerprintless''. In our design, we tried to avoid
obvious distinguishers and minimised the interaction surface for attackers who lack the shared
secret. But since \pt{} does not mimic a cover protocol, it hides within the set of \emph{unknown
rather than known} protocols. As long as a censor's network features a high diversity of
network protocols, the censor is unable to fully control or model, we expect our approach to provide
a decent level of protection.

% % % % % % % % % % % % % % % % % % % % % % % % % % % % % % % % % % % % % % % % % % % % % % % % % %
\section{Conclusion}
\label{sec:conclusion}
% % % % % % % % % % % % % % % % % % % % % % % % % % % % % % % % % % % % % % % % % % % % % % % % % %
We presented \pt{}; a transport protocol which provides lightweight obfuscation for application
protocols such as Tor and VPN. The two major contributions of our protocol are the ability to defend
against \emph{active probing} and \emph{protocol classifiers}. We achieve the former by proposing
two authentication mechanisms---one general-purpose and the other specifically for Tor---and the
latter by proposing morphing techniques to disguise packet lengths and inter arrival times.

We further developed a prototype of \pt{} and used it to conduct an experimental evaluation. In
particular, we discussed the effectiveness of our obfuscation techniques as well as \pt{}'s
overhead. Our evaluation suggests that \pt{} can provide strong protection against censors who do
not overblock significantly. Finally, we believe that the low protocol overhead makes \pt{}
comfortable to use for web surfing and other low-latency applications.

% % % % % % % % % % % % % % % % % % % % % % % % % % % % % % % % % % % % % % % % % % % % % % % % % %
\section*{Acknowledgements}
% % % % % % % % % % % % % % % % % % % % % % % % % % % % % % % % % % % % % % % % % % % % % % % % % %
We want to thank George Kadianakis, Harald Lampesberger, Stefan Lindskog, and Michael Rogers who all
provided valuable feedback which improved this paper. We further want to thank Internetfonden
of the Swedish Internet Infrastructure Foundation for supporting the main author's work with a
research grant.

Our code is publicly available at
\url{http://www.cs.kau.se/philwint/scramblesuit/}. Finally, we point out that all of the references
listed below contain a link to an open access version of the respective resource. Please consider
doing the same in your papers.

\printbibliography

\end{document}